\documentclass[a4paper,aps,prl,amsmath,amssymb,showpacs,twocolumn,10pt]{revtex4-1}
\usepackage{epsfig}
\usepackage{graphicx}
\usepackage{amsmath,amssymb,psfrag,textcomp}
\usepackage{amsthm,upgreek,bm}

\newcommand{\be}{\begin{equation}}
\newcommand{\ee}{\end{equation}}
\newcommand{\bd}{\begin{displaymath}}
\newcommand{\ed}{\end{displaymath}}

\begin{document}
\title{Large Deviations of the Smallest Eigenvalue of the Wishart-Laguerre Ensemble}
\author{Eytan Katzav}
\author{Isaac P\'erez Castillo}
\address{Department of Mathematics, King's College London, Strand, London WC2R 2LS, United Kingdom}
\begin{abstract}
We consider the large deviations of the smallest eigenvalue of the Wishart-Laguerre Ensemble. Using the Coulomb gas picture we obtain rate functions for the large fluctuations  to the left and the right of the hard edge. Our findings are compared with known exact results for $\beta=1$ finding good agreement. We also consider the case of almost square matrices finding new universal rate functions describing large fluctuations.
\end{abstract}
\maketitle
\begin{figure}[ht]
\centerline{\includegraphics[width=8.5cm,height=5.4cm]{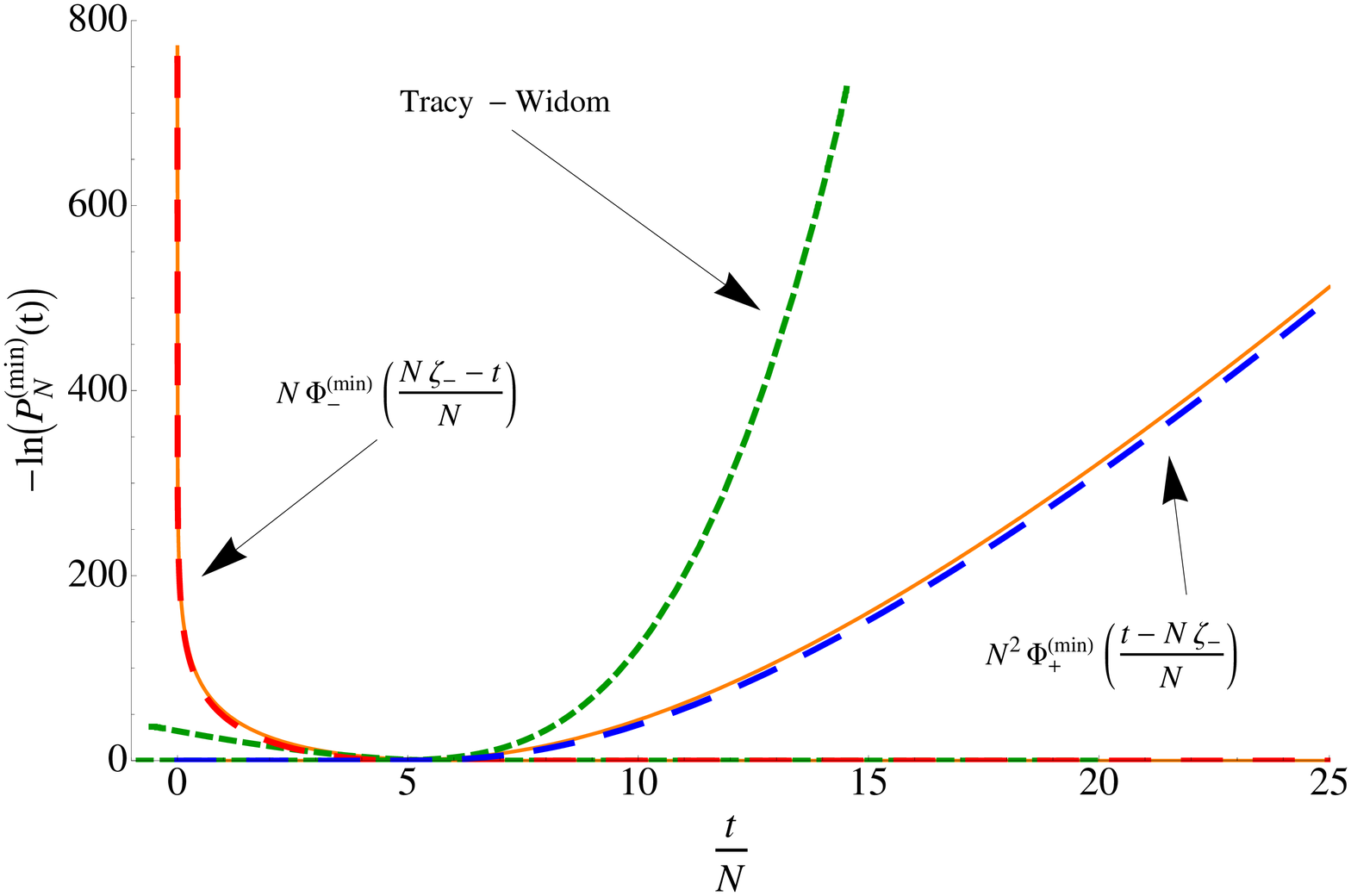}}
 \caption{(color online). Results for the smallest eigenvalue distribution $-\ln P^{(\text{min})}_N(t)$ vs. the scaled variable $t/N$. Here, $N = 11$ and $M=110$ ($c=1/10$), Wishart matrices are real ($\beta=1$). The large deviation functions (dashed lines) compare very well with the exact result of \cite{Edelman1991} (solid line), while TW (dotted line) describes only small fluctuations.}
 \label{fig:exact}
\end{figure}

In the early $50$s, when not much was known about the intricacies of complex atomic nuclei, Wigner suggested to replace the underlying physics of the problem by its apparent statistical features \cite{Wigner1951}. Surprisingly, it turned out that such statistical approach was more useful than anyone could have anticipated,  enabling him to work  out the nuclei level spacing distribution. Random Matrix Theory (RMT) has played a central role in various branches of science, since its first appearance in Statistics by Wishart in 1928 \cite{Wishart1928}, through QCD \cite{Akemann2000}, random graphs \cite{Rogers2008}, wireless communications \cite{Tulino} and computational biology \cite{Eisen1998} - to mention a few. And although through more than half a century different, seemingly unrelated, problems have been linked via RMT, certain questions about eigenvalue distributions are still poorly explored. One such question is the distribution of the smallest eigenvalue.\\
In physics, for instance, classical disordered systems offer the ideal context where RMT concepts and tools may be applied. Here, complicated systems are simplified by  using a random Hamiltonian, where the smallest eigenvalue is of special interest as it is associated with the ground state. In quantum entanglement, the smallest eigenvalue is a useful measure of entanglement \cite{Nadal2010}.\\
In mathematics, the minimal eigenvalue appears naturally in many contexts, such as the study of the geometry of random polytopes \cite{Litvak2005}. It is also extremely relevant for the question of invertibility of random matrices \cite{Rudelson2008}, and recently played an important role in the exploding field of compressive sensing \cite{Candes2006}, where fluctuations of the minimal eigenvalue set the bounds on the number of random measurements needed to fully recover a sparse signal.\\
In statistics, a very important technique used to detect hidden patterns in complex, high-dimensional datasets is called "Principal Components Analysis" (PCA). The idea is to take a data matrix $\bm{X}$, and to transform its covariance matrix $\bm{W} = \bm{X}^\dag \bm{X}$ into a new coordinate system such that the greatest variance by any projection of the data comes to lie on the first coordinate. Technically speaking, one identifies eigenvalues and eigenvectors of $\bm{W}$, and ignores the components corresponding to the lowest eigenvalues, as these eigenmodes contain the least important information. The smallest eigenvalue of $\bm{W}$ determines the largest eigenvalue of $\bm{W}^{-1}$, and is important in Hotelling's T-square distribution \cite{Muirhead1982}, for example.\\
The purpose of this work is to provide a simple physical method, based on the Coulomb gas method in statistical physics \cite{Dean2006,*Dean2008}, that allows us to compute analytically the probability of the smallest eigenvalue in the Wishart-Laguerre ensemble.\\
We consider an ensemble $G(M,N)$ of $M\times N$ rectangular matrices $\bm{X}\in G(M,N)$, which  are drawn from a Gaussian distribution $P(\bm{X}) \propto \exp\left[-{\textstyle{\beta \over 2}}\text{Tr}\left(\bm{X}^\dag \bm{X}\right)\right]$, where $\beta$ is the Dyson index with classical values $\beta=1,2$, and $4$,  corresponding to the real, complex and quaternionic cases respectively.  The Wishart ensemble $W(M,N)$ is defined as the set of $N\times N$ matrices $\bm{W}=\bm{X}^{\dag}\bm{X}$. If $\bm{\lambda}$ denotes the $N$ eigenvalues of a Wishart matrix, their joint PDF reads
\begin{equation}
\begin{split}
P(\bm{\lambda})=\frac{1}{Z(0)}e^{-\beta F(\bm{\lambda})/2}
\label{eq:jed}
\end{split}
\end{equation}
with $Z(0)$ a normalisation constant and $F(\bm{\lambda})$ defined as:
\begin{equation}
 F(\bm{\lambda})=\sum_{i=1}^N\lambda_{i}-u\sum_{i=1}^N\ln(\lambda_i)-\sum_{i\neq j}\ln |\lambda_i-\lambda_{j}|
\end{equation}
where $u=1+M-N-\frac{2}{\beta}$.\\
We will restrict ourselves to the case $M\geq N$. It is well known \cite{MP67} that for large $N$ the density of eigenvalues $\rho_{N}(\lambda)=\frac{1}{N}f(\frac{\lambda}{N})$ is given by the Mar\v{c}enko-Pastur law:
\begin{equation}
f(x)=\frac{1}{2\pi x}\sqrt{(x-\zeta_{-})(\zeta_{+}-x)}\,\openone_{x\in[\zeta_{-},\zeta_{+}]}
\label{eq:1}
\end{equation}
with $\zeta_{\pm} =(1\pm\sqrt{1+\alpha})^2$, $\alpha=\frac{1-c}{c}$ and $c=\frac{N}{M}$. The indicator $\openone_{x\in D}$ takes the value $1$ if $x\in D$ and $0$ otherwise. The points $\zeta_{-}$ and $\zeta_{+}$ are usually called the hard edge and the soft edge of the distribution \eqref{eq:1}, respectively. The case $\alpha=0$ (or $M=N$) corresponds to square matrices, and the case $\alpha=\mathcal{O}(1/N)$ (or $M-N=\mathcal{O}(1)$) is referred to here as "almost square matrices".\\
We will focus on the statistical fluctuations around the hard edge, which are captured by the probability distribution of the smallest eigenvalue $\rho^{(\text{min})}_{N}(\lambda)$, from which any other statistical property of the hard edge may be inferred. A classical result \cite{Silverstein1985} shows that $\lambda_{\text{min}}$ converges almost surely to $\zeta_-$ as $N \rightarrow \infty$. However, the concentration of the minimum around this value is generally unknown, and can be a challenging task.\\
The distribution of the smallest eigenvalue $\rho^{(\text{min})}_{N}(\lambda)$ (as well as for the maximal eigenvalue) has been formally expressed using Multivariate Hypergeometric Functions and Zonal Polynomials. This was first done long ago for real matrices ($\beta=1$) in \cite{Krishnaiah1971}, then generalised to complex matrices ($\beta=2$) in \cite{Ratnarajah2004} and only recently to any $\beta$ in \cite{Dumitriu2008}. These expressions are  rigorous but often not easy to evaluate and manipulate (although a real breakthrough has occurred recently with a new algorithm that can calculate such quantities with complexity that is only linear in the size of the matrix \cite{Koev2006}, as well as available dedicated packages \cite{Dumitriu2007}).\\
Another line of explicit expressions involve a determinantal representation of the distribution of the smallest eigenvalue (as well as other order statistics) for finite $N$ and $M$ (see \cite{Akemann2008} and references therein). These expression are not always easy to implement, and expecially when $M-N$ is large.\\
More explicit expressions have been derived by Edelman in \cite{Edelman1988,Edelman1991} for $\beta=1$ (for $\beta=2$ as well, but only for square matrices). These expressions require the knowledge of some polynomials (different ones for any $M$ and $N$). While manageable and useful for small matrices, this turns out to be impractical for large matrices, since no explicit formula exists for these polynomials. It is especially changeling to extract other relevant statistical properties around the hard edge like, for instance, the distribution of the typical (order $\mathcal{O}(N^{p})$ with $0<p\leq 1$) or large (order $\mathcal{O}(N)$) fluctuations around their mean values.\\
The typical fluctuations of the smallest eigenvalue have been recently studied rigorously in \cite{Feldheim2010}, where it is shown that the smallest eigenvalue $\lambda_{\text{min}}$ follows the Tracy-Widom (TW) distribution, that is, the typical fluctuations of $\lambda_{\text{min}}$ can be expressed as
\begin{equation}
\begin{split}
\lambda_{\text{min}}=\zeta_{-}N-\zeta_{-}^{2/3}c^{1/6} N^{1/3}\chi
\end{split}
\end{equation}
where $\chi$ follows a TW distribution $g_{\beta}(\chi)$ \cite{Feldheim2010}. This result was proven for the case $c<1$ in the large $N$-limit, and so strictly speaking, it does not apply to either square or almost square matrices.\\
The knowledge about large deviations of the smallest eigenvalue is, as far as we are aware of, mainly unexplored. We would like here to correct the situation. To do so we will use the Coulomb gas approach \cite{Dean2006,Dean2008}. Starting with Eq.~\eqref{eq:jed} the cumulative probability of the minimum $P^{(\text{min})}_{N}(t) \equiv P(\lambda_{\text{min}}\geq t)$ can be written as
\begin{equation}
P^{(\text{min})}_{N}(t) =\int_{t}^{\infty } d\lambda\rho^{(\text{min})}_{N}(\lambda)=\frac{Z(t)}{Z(0)}
\label{eq:prob}
 \end{equation}
with
\begin{equation}
 Z(t)=\int_t^\infty \cdots \int_t^\infty
d\lambda_1\cdots d\lambda_N\, e^{-\beta F(\bm{\lambda})/2}
\label{eq:Zl}
\end{equation}
In this framework $Z(t)$ is understood as the partition function of a $2D$ Coulomb gas of charged particles restricted on a $1D$ line, in an external  linear-log potential and a hard wall at $t$. The idea, as in \cite{Dean2008},  is to evaluate $Z(t)$ by using the saddle-point approximation. It is important  to notice that the expression \eqref{eq:prob} is exact and that while the  saddle-point method provides the exact density of eigenvalues for large $N$,  it is only able to capture the large deviations to the right of the smallest eigenvalue from $\rho^{(\text{min})}_{N}(\lambda)$. This is not entirely surprising as $\rho_{N}(\lambda)$ is a collective quantity while $\rho^{(\text{min})}_{N}(\lambda)$ is not.\\
The calculation goes along similar lines as in \cite{Vivo2007,Chen1994}, so we shortly describe the needed steps. To apply the saddle-point method to \eqref{eq:Zl}, one first introduces the function $\varrho(\lambda)=\frac{1}{N}\sum_{i=1}^N\delta\left(\lambda-\lambda_i\right)$. This allows us to write the $Z(t)$ as a path integral over $\varrho(\lambda)$ and its Lagrange multiplier $\widehat{\varrho}(\lambda)$.  Minimising the corresponding functional with respect to $\widehat{\varrho}(\lambda)$,  allows to eliminate this multiplier and to unveil the meaning of $\varrho(\lambda)$ as a constrained spectral density. After rescaling the eigenvalues ($\lambda=x N$ and $\zeta=t/N$), $Z(t)$ takes the form
\begin{equation}
\begin{split}
Z(t)&=\int\{\mathcal{D} f\} e^{-\frac{\beta}{2}N^2 \mathcal{S}[f(x)]}
\end{split}
\end{equation}
with the rescaled density $\varrho(\lambda)=\frac{1}{N}f(\frac{\lambda}{N})$, and the action
\begin{eqnarray}
\mathcal{S}[f(x)]&=& \int_{\zeta}^{\infty} d x f(x) x- \left(\alpha+{\textstyle{{\beta  - 2} \over {\beta N}}}\right)\int_{\zeta}^{\infty} dxf(x)\ln x \nonumber\\
&-&\int_{\zeta}^{\infty}\int_{\zeta}^{\infty} dx dx'f(x)f(x')\ln |x-x'|\label{eq:action}\\
&+&{\textstyle{2 \over {\beta N}}}\int_{\zeta}^{\infty} d xf(x)\ln f(x)+C_1\left[\int_{\zeta}^{\infty}dxf(x)-1 \right] \nonumber
\end{eqnarray}
As we will apply the saddle-point method we neglect the terms which are of order $\mathcal{O}(N^{-1})$ in the preceding expression for $\mathcal{S}[f(x)]$. Note however that while the first term in the subleading correction can be easily kept, the entropic term makes the evaluation of the saddle-point equation a changeling task. The next step is to look for the saddle point of the action. The saddle point equation is basically $\delta\mathcal{S}[f(x)]/\delta f(x) = 0$. It is useful to differentiate the saddle-point equation once with respect to $x$:\begin{equation}
\begin{split}
\frac{1}{2}-\frac{\alpha}{2x}=P\int_\zeta^\infty\frac{f_\star(x)}{x-x'}dx'\,,\quad x \in [\zeta,\infty)
\end{split}
\end{equation}
where $f_\star(x)$ is the value of $f(x)$ at the saddle-point. This is a Tricomi integral equation, and is solved as in \cite{Vivo2007}, so we report the final result: For $\zeta\in[0,\zeta_{-}]$, $f_\star(x)$ is given by Eq.~(\ref{eq:1}), while for $\zeta\in[\zeta_{-},\infty)$,
\begin{equation}
f_\star(x)=\frac{\sqrt{U-x}}{2 \pi\sqrt{x-\zeta}} \left( \frac{x-\alpha\sqrt{\zeta/U}}{x} \right) \,\openone_{x\in[\zeta, U]}
\end{equation}
with $U \equiv U(c,\zeta)=w^2(c,\zeta)$ and  $w(c,\zeta)$ given by
\begin{equation}
w(c,\zeta)=\frac{2p}{ 3\rho^{1/3}}\cos\left(\frac{\theta+2 \pi }{3}\right) \, ,
\end{equation}
with $p=-[\zeta+2(\alpha+2)]$, $q=2\alpha\sqrt{\zeta}$, $\rho=\sqrt{-\frac{p^3}{27}}$, $\theta=\text{atan}\left(\frac{2\sqrt{B}}{q}\right)$ and $B=-\left(\frac{p^3}{27}+\frac{q^2}{4}\right)$. Using these results and after a long and tedious calculation we obtain the following results for the distribution: for $t\in [0,N\zeta_{-}]$ we have $P^{(\text{min})}_{N}(t)=1$, and 
\begin{equation}
P^{(\text{min})}_{N}(t)=e^{-\beta N^2\Phi^{(\text{min})}_{+}\left(\frac{t-N\zeta_{-}}{N}\right)} \,,\,\,  t\in[N\zeta_{-},\infty)
\label{eq:right}
\end{equation}
with the right rate function $\Phi^{(\text{min})}_{+}(x)$ being
\begin{equation}
\begin{split}
\Phi^{(\text{min})}_{+}(x)&= \frac{1}{2}\left[S\left(x+\zeta_-\right)-S\left(\zeta_{-}\right)\right], \,\, x\in [0,\infty)\\
\end{split}
\end{equation}
and the action $\mathcal{S}(\zeta) \equiv \mathcal{S}[f_\star(x)]$ given by
\begin{eqnarray}
S\left(\zeta\right)&=&\frac{\zeta+U}{2}-\frac{\Delta^2}{32}-\ln\left(\frac{\Delta}{4}\right)+\frac{\alpha}{4}\left(\sqrt{U} -\sqrt{\zeta}\right)^2 \nonumber\\
&+&\frac{\alpha^2}{4}\ln\left(\zeta U\right)-\alpha(\alpha+2)\ln\left[\frac{\sqrt{U} +\sqrt{\zeta}}{2}\right] \, ,
\end{eqnarray}
where, $\Delta=U(c,\zeta)-\zeta$. As mentioned above, the saddle-point approximation is only able to capture the large fluctuations to the right of $\lambda_{\text{min}}$ as it has a hard wall on the left side \eqref{eq:right}. Fortunately, the authors of \cite{Majumdar2009} came up with a beautiful physical argument to overcome this shortcoming and to estimate in their case the large deviations from the maximal eigenvalue. Applied to the fluctuation to the left of $\lambda_{\text{min}}$ this method yields
\begin{equation}
\begin{split}
P^{(\text{min})}_{N}(t)&\sim e^{-\beta N\Phi^{(\text{min})}_{-}\left(\frac{N\zeta_{-} -t}{N}\right)}\,,\quad  t\in[0,N\zeta_{-}]
\end{split}
\end{equation}
with the left rate function $\Phi^{(\text{min})}_{-}(x)$ given by:
\begin{eqnarray}
&&\Phi^{(\text{min})}_{-}(x)=- {\textstyle{\alpha  \over 2}}\ln
\left( {1 - {\textstyle{x \over {{\zeta _ - }}}}} \right) - {\textstyle{1 \over 2}}\sqrt x \sqrt {x + \Delta_{-}} \\
&&+ 2\ln \left( {{\textstyle{{\sqrt {x + {\Delta_- }}  - \sqrt x } \over {\sqrt {{\Delta_- }} }}}} \right) + \alpha \ln \left( {1 +
2\sqrt {\textstyle{x \over {\zeta_-}}} {\textstyle{{\sqrt {x + \Delta_-} - \sqrt x } \over {\Delta_-}}}} \right) \nonumber
\end{eqnarray}
with $x\in[0,\zeta_{-}]$ and $\Delta_-=\zeta_+ - \zeta_- = 4\sqrt{\alpha+1}$.\\
We now check that the smallest large fluctuations predicted by our results match the largest typical fluctuations given by the TW distribution \cite{Feldheim2010}. Expanding the rate functions $\Phi^{(\text{min})}_{\pm}(x)$ around $x=0$ gives:
\begin{eqnarray}
\Phi^{(\text{min})}_{-}(x)&\underset{x\to0}{\sim}\frac{2}{3 \zeta_{-} c^{1/4}} x^{3/2} \\
 \Phi^{(\text{min})}_{+}(x)&\underset{x\to0}{\sim}\frac{1}{24\zeta_{-}^2\sqrt{c}} x^3
\end{eqnarray}
which yields the following expression of $P^{(\text{min})}_{N}(t)$ for the smallest large fluctuations of $\lambda_{\text{min}}$:
\begin{equation}
P^{(\text{min})}_{N}(t)\sim\left\{\begin{split}
&\exp\left(- {\textstyle{{2\beta } \over 3}} \chi^{3/2}(t)\right),\quad t\in [0,N\zeta_{-}] \\
&\exp\left(- {\textstyle{{\beta } \over 24}} |\chi(t)|^3\right)  ,\quad t\in [N\zeta_{-},\infty)
\end{split}\right. \nonumber
\end{equation}
with $\chi(t) \equiv -(N\zeta_{-}-t)/(N^{1/3}\zeta_{-}^{2/3}c^{1/6})$. This result obviously agrees with TW distribution for large $|\chi|$ \cite{Feldheim2010,Tracy1994,*Tracy1996}.\\
For larger atypical fluctuations, we have the following asymptotic behaviours of the rate functions:
\begin{eqnarray}
 \Phi^{(\text{min})}_{-}(x)&&\underset{x\to \zeta_{-}} {\sim}\left[ {-{\textstyle{\zeta_{-} \over 2}} - \sqrt {\zeta_{-}}  + {\textstyle{\ln(c) \over {2c}}} + \alpha \ln \alpha } \right] \\
 &&-{\textstyle{\alpha  \over 2}}\ln \left( {\zeta_{-} - x} \right) + \mathcal{O}\left( {\zeta_{-} - x} \right) \, , \nonumber \\
 \Phi^{(\text{min})}_{+}(x)&&\underset{x\to \infty} {\sim} {\textstyle{1 \over 2}}x - {\textstyle{\alpha  \over 2}}\ln x + \left[ {{\textstyle{\zeta_{-} \over 2}} + {\textstyle{3 \over 4}} - {\textstyle{S(\zeta_{-}) \over 2}}} \right] .
\end{eqnarray}
Interestingly, these results can be compared with the exact asymptotic behaviours predicted in \cite{Edelman1991} for the case $\beta=1$ and large $N$. It turns out that for $\Phi^{(\text{min})}_{-}(x)$ the leading logarithmic behaviour agrees perfectly with \cite{Edelman1991}, while the constant term cannot be rigorously compared, since it is not available for any $N$ in the exact treatment \cite{Edelman1991} (although we know that a constant term exists). For $\Phi^{(\text{min})}_{+}(x)$ we again see perfect agreement with the linear and logarithmic terms, but the constant deviates from the exact one given in \cite{Edelman1991}.\\
A comparison of these results, as well as with results of simulations and the TW distribution \footnote{We have used the table provided by Pr{\"a}hofer and Spohn at http://www-m5.ma.tum.de/KPZ/} are summarized in Fig.~\ref{fig:exact}. Note that  exact results are only available for $\beta=1$ \cite{Edelman1991}, and so a comparison to exact results for ensembles other than the real Wishart case is not possible.\\

\begin{figure}[ht]
\centerline{\includegraphics[width=8.5cm,height=5.4cm]{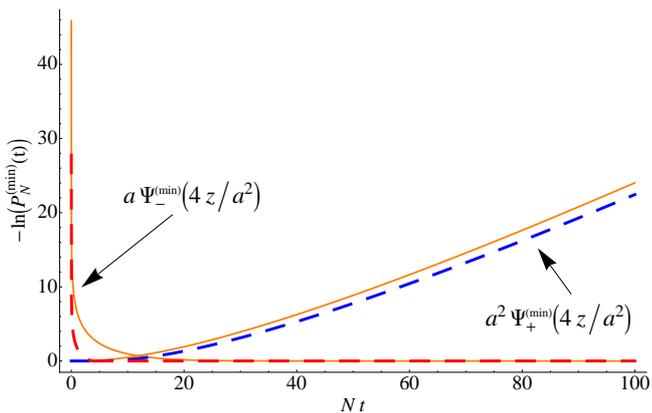}}
 \caption{(color online). Results for the smallest eigenvalue distribution $-\ln P^{(\text{min})}_N(t)$ vs. the scaled variable $t N$. Here, $N = 200$ and $a=5$, Wishart matrices are real ($\beta=1$). The large deviation functions derived here (dashed lines) compare well with the exact result of \cite{Edelman1991} (solid line).}
 \label{fig:almost-square}
\end{figure}

\noindent We now pay special attention to the case of almost square matrices, namely when $M=N+a$ with $a$ an integer of order unity (i.e. $\alpha=a/N$, so $\alpha$ is of order $1/N$). Here it is useful to use the scaling $z=N t=N^2\zeta$ to unravel non-trivial results. Using the Coulomb gas approach we find that in this special case, the PDF of $\lambda_{\text{min}}$ does not simply approach a delta function, in the large $N$ limit, as in \cite{Silverstein1985}. Instead, the cumulative distribution of $z$ has a $N$-independent limiting shape as shown in Fig.~\ref{fig:almost-square} for $\beta=1$, and whose large fluctuations are described by the Coulomb gas prediction:
\begin{equation}
 P^{(\text{min})}_N(z) \sim \left\{ \begin{array}{l}
 \exp \left( { - \beta a{\Psi^{(\text{min})}_{-} }\left( {{\textstyle{{4z} \over {{a^2}}}}} \right)} \right) \,, \quad z \in [0,a^2/4] \\
 \exp \left( { - \beta {a^2}{\Psi^{(\text{min})}_{+} }\left( {{\textstyle{{4z} \over {{a^2}}}}} \right)} \right) \,, \quad z \in [a^2/4,\infty)\\
 \end{array} \right.
 \label{eq:almost-square}
\end{equation}
with $\Psi^{(\text{min})}_{+} (x) = \left( {x - 4\sqrt x  + 3 + \ln x} \right)/8$ and $\Psi^{(\text{min})}_{-} (x) = \ln {\textstyle{{1 + \sqrt {1 - x} } \over {\sqrt x }}} - \sqrt {1 - x}$. The functions $\Psi^{(\text{min})}_{\pm}(x)$ are universal, in the sense that  they are $N,\beta$ and $a$-independent. Note that in this particular regime one needs to keep the first $1/N$ term appearing  \eqref{eq:action} as it is of the same order as $\alpha=a/N$. This can be accounted for easily by replacing in the preceding expressions $a\to a(\beta)=a+(\beta-2)/\beta$, as long as $a(\beta)$ is non-negative. To our knowledge, this is the first time that this case is discussed and shown to be universal (even numerically),  and obviously no general explicit predictions for its shape have been proposed, apart of the particular case of   $a(\beta)=0$ for which Eq. \eqref{eq:almost-square} yields $P^{(\text{min})}_{N}(z)=e^{-\beta z/2}$ in agreement with the exact result of \cite{Edelman1991} (for $a=\beta=1$) and \cite{Edelman1988} (for $a=0$, $\beta=2$). Note also that the results for TW distribution reported in \cite{Feldheim2010} do not formally apply to almost square matrices.\\
In  Fig.~\ref{fig:almost-square} we have compared our finding with Edelman's exact result for $\beta=1$, $N=200$ and $M=205$.

\noindent It is important to point out that while the results in \cite{Edelman1991} are exact, they become difficult to evaluate for large values of $N$ and a daunting task to extract exact results about large fluctuations for large matrices. In contrast, the result \eqref{eq:almost-square} provide information of large fluctuations for any value of $\beta$ and $a$, when $a(\beta)\geq0$.\\

To summarise, in this work we study the large deviation functions
of the smallest eigenvalue of the Wishart-Laguerre Ensemble using the
Coulomb gas approach. We obtain explicit expressions for both the
right and left rate functions for general $\alpha$. We also
highlight the existence of a special regime for almost square
matrices, where an interesting limit distribution exists, described
by universal rate functions $\Psi^{(\text{min})}_{\pm}(x)$. We were
able to provide predictions for $a(\beta) \ge 0$, which leaves the
question of $a(\beta) < 0$ open for further research. Another
interesting open question is regarding the typical fluctuations for
the smallest eigenvalue for square and almost square matrices, which
are not captured by the TW distribution.

\bibliography{bio}

\end{document}